
********

\magnification\magstep 1
\vsize=8.5 true in
\hsize=6 true in
\voffset=0.5 true cm
\hoffset=1 true cm
\baselineskip=20pt

\bigskip
\bigskip
\bigskip
\vskip 5 truecm
\centerline{\bf EFFECTS OF DISORDER IN A DILUTE BOSE GAS}

\vskip 1.5 truecm

\centerline{S. Giorgini, L. Pitaevskii$^*$ and S. Stringari}
\bigskip
\centerline{{\it Dipartimento di Fisica, Universit\'a di Trento,
  I-38050 Povo, Italy}}
\vskip 2 truecm
{\it{\bf Abstract.} We discuss the effects of a weak
random external potential
on the properties of
the dilute Bose gas at zero temperature.
The results recently obtained by Huang and Meng
for the depletion of the condensate and of the
superfluid density are recovered. Results for the shift of
the velocity of sound  as well as for its damping due to collisions
with the external field are presented. The damping of phonons is calculated
also for dense superfluids.}
\vskip 5.0truecm

\par\noindent Preprint UTF 253, December 1993

\bigskip
\bigskip

\par\noindent PACS Numbers: 67.40.-w, 67.40.Db
\vskip 1.5truecm

\par\noindent $^*$ Permanent address: Kapitza Institute for Physical Problems,
ul. Kosygina 2, 117334 Moscow.

\vfill\eject

\par\noindent
{\bf  1. INTRODUCTION}
\bigskip

The role of disorder in Bose systems has been the object of several theoretical
works in the last years$^{1-7}$. In this context an
important issue is to understand the effects of
disorder on the behaviour of the dilute Bose gas (DBG) where one expects to
obtain analytic results at least for weak disorder. This problem was recently
considered by the authors of Ref. [3] who investigated the consequences of
a random external potential on the depletion of the condensate and of the
superfluid component of
the system.
A striking result emerging from this analysis is that disorder
is more active in reducing superfluidity than in depleting
the condensate.

The main purpose of the present work is to investigate
the consequences of disorder on the propagation of phonons.
In particular we study the shift of the sound velocity
as well as its damping
generated by scattering with the random
external potential.

Our formalism is based on dispersive quantum hydrodynamics at zero temperature.
This formalism not only accounts in a natural way
for the behaviour of the dilute
Bose gas, but can be employed, in the hydrodynamic regime, to predict
important properties of dense
superfluids such as liquid $^4$He.

The paper is organized as follows: in Section 2 we develop the formalism of
quantum hydrodynamics in the presence of random external potentials. In Sect. 3
we recover, in the limit of weak disorder, the results of ref.[3]
for the depletion of the condensate and of the superfluid density.
In Sect.4 we explore the consequences of disorder on the dynamic structure
factor. In particular we calculate the change in the velocity of sound and the
damping of phonons induced by disorder. The expression for the damping is
also extended to the case of strongly interacting superfluids.

\bigskip
\par\noindent
{\bf 2. FORMALISM OF QUANTUM HYDRODYNAMICS.}
\bigskip

Let us consider a uniform superfluid system at $T=0$. Our approach to
the problem starts from a macroscopic description:
we write the energy functional of the system in terms of the superfluid
velocity, ${\bf v}_s$, and the variation of the density with respect to the
equilibrium value, $\rho'=\rho-\rho_0$. We find$^8$
$$
E = {\rho_0\over2}\int d^3r\;v_s^2({\bf r})+{1\over2}\int d^3r \int
d^3r' \chi^{-1}(|{\bf r}-{\bf r'}|)\rho'({\bf r})\rho'({\bf r'})
$$
$$
\;\;\;\;\;\;\;\;\;\;
+{1\over2}\int d^3r\;\rho'({\bf r})v_s^2({\bf r})
\eqno(1)
$$

The superfluid behaviour of the system fixes the irrotational law for the
velocity field: ${\bf v}_s=\nabla\varphi$.
In terms of the Fourier components of the density and of the velocity
potential,
the energy
functional can be written as the sum of a quadratic and a cubic term in
the fluctuations
$\varphi_{{\bf k}}$ and $\rho_{{\bf k}}'$: $E=H_0+H_1$, where
$$
H_0 = {\rho_0\over2}\sum_{\bf k}k^2|\varphi_{\bf k}|^2 +
{1\over2}\sum_{\bf k}\chi^{-1}(k)|\rho'_{\bf k}|^2
\eqno(2)
$$
and
$$
H_1 = {1\over2\sqrt{V}}\sum_{{\bf k},{\bf k}'}
{\bf k}\cdot{\bf k}'\;\varphi_{{\bf k}}\rho_{{\bf k}'-{\bf k}}'
\varphi_{-{\bf k}'}
\eqno(3)
$$
The quantity $\chi^{-1}(k)$ is the Fourier transform of $\chi^{-1}(r)$.
Terms of third and higher order in $\rho'$ have not been included in eqs.
(1), (3) since in the DBG limit they do not contribute to the leading
order. Applicability of eq. (1) is limited to the region of low temperatures
where the whole system is superfluid.

The quadratic functional of equation (2) gives the expression for the energy in
the framework of the "harmonic dispersive hydrodynamics", where the
compressibility of the system, given by $\chi(k)$, is momentum dependent.
In the long wavelength limit the
hydrodynamic hamiltonian (2) coincides with the usual phonon hamiltonian with
the velocity of sound given by $c=(\rho_0/\chi(0))^{1/2}$.

Let us neglect for the moment the anharmonic term $H_1$ and consider only the
harmonic hamiltonian $H_0$.
By quantizing the fields $\varphi$ and $\rho'$ we can cast the hamiltonian into
the diagonal form
$$
\hat{H}_0 = \sum_{\bf k}\omega_k\hat{c}_{\bf k}^{\dagger}
\hat{c}_{\bf k}
\eqno(4)
$$
where
$\hat{c}_{\bf k}^{\dagger}$ ($\hat{c}_{\bf k}$) are the creation
(annihilation) operators relative to an elementary excitation carrying impulse
${\bf k}$ and we have ignored the constant term due to the zero point motion.
Equation (4) shows that the system can
be described as a gas of non-interacting elementary excitations with energies
$$
\omega_k = \sqrt{{\rho_0 k^2\over \chi(k)}} .
\eqno(5)
$$

The quasiparticle operators $\hat{c}_{\bf k}^\dagger,\hat{c}_{\bf
k}$ are related to the operators $\hat{\varphi}_{\bf k}$ and $\hat{\rho}'_{\bf
k}$ through the following expressions
$$
\delta\hat{\varphi}_{\bf k}=\hat{\varphi}_{\bf k} -
\langle\hat{\varphi}_{\bf k}\rangle =
({\omega_k\over 2\rho_0 k^2})^{1/2}\;(\hat{c}_{\bf k}+
\hat{c}_{-{\bf k}}^{\dagger})
$$
$$
\delta\hat{\rho}_{\bf k}=\hat{\rho}_{\bf k}' -
\langle\hat{\rho}_{\bf k}'\rangle =
i\;({\rho_0 k^2\over2\omega_k})^{1/2}\;(\hat{c}_{\bf k}-
\hat{c}_{-{\bf k}}^{\dagger})
\eqno(6)
$$
where we have introduced the fluctuation operators $\delta\hat{\varphi}_{\bf
k}$ and $\delta\hat{\rho}_{\bf k}$ which, in the absence of disorder, coincide
with the operators $\hat{\varphi}_{\bf k}$ and $\hat{\rho}_{\bf k}'$.

In the dilute Bose gas limit the frequencies $\omega_k$ are given by the most
famous Bogoliubov spectrum
$$
\omega_k = {k\over2m}(k^2+4\rho_0v_0)^{1/2}
\eqno(7)
$$
and the compressibility $\chi(k)$ takes the form
$$
\chi(k) = {4\rho_0 m^2 \over k^2 + 4\rho_0 v_0} \, .
\eqno(8)
$$
In terms of the scattering length  $a$
the interparticle potential $v_0$ is written as: $v_0=4\pi a/m$.

Let us now add a perturbative field in the form of a random external potential
$U({\bf r})$ coupled to the density through the interaction hamiltonian
$$
\hat{V} = {1\over\sqrt{V}}\sum_{\bf k}U_{\bf k}\hat{\rho}'_{-{\bf k}}
\eqno(9)
$$
In eq. (9) $U_{\bf k}$ is the Fourier transform of the external potential.
In order to investigate in a
simple way how the random potential affects the behaviour of the system, we
will often make the white noise assumption in which the external potential
is described by a single parameter $R_0$ [3]:
$$
{1\over V}\langle|U_{\bf k}|^2\rangle = {R_0\over m^2}
\eqno(10)
$$
and where the average is assumed to be  of Gaussian type.
In eq. (10) $m$ is the mass of the particles and the dimension of $R_0$
is consequently (energy)$^2$ $\times$ (length)$^3$.

In the presence of weak disorder the expectation value of the density operator
on the ground state can be evaluated by means of linear response theory:
$$
\langle\hat{\rho}_{\bf k}'\rangle=-{1\over\sqrt{V}}U_{\bf k}\;\chi(k)
\eqno(11)
$$
whereas the expectation value of the velocity potential operator
is not affected by the external static field:
$\langle\hat{\varphi}_{\bf k}\rangle=0$.
By writing both the hamiltonians $\hat{H}_0$ and $\hat{V}$
in terms of the operators $\hat{c}_{\bf k}$, $\hat{c}_{\bf k}^{\dagger}$ we get
the result
$$
\hat{H} = -{\rho_0\over2}\sum_{\bf k}{1\over V}\langle|U_{\bf k}|^2\rangle
{k^2\over\omega_k^2}
+\sum_{\bf k}\omega_k\hat{c}_{\bf k}^\dagger
\hat{c}_{\bf k}
\eqno(12)
$$
for the harmonic hamiltonian $\hat{H}=\hat{H}_0+\hat{V}$.
The first term in (12) gives the correction to the ground state energy due to
the external field.
This quantity can be calculated also beyond the dilute Bose gas approximation.
In fact by applying linear reponse theory (eq.(11)) to the external field
(9) one finds the general result
$$
\delta E = -{1\over 2} \sum_{\bf k}{1\over V}\langle|U_{\bf k}|^2\rangle\chi(k)
\eqno(13)
$$
holding also for strongly interacting systems.
Equation (13) reduces to the first term in the
right hand side of eq.(12) in the Bogoliubov limit where eq.(5) holds.
It is important to notice that if one makes the white noise assumption (10),
the correction to the ground state energy (13) contains an ultraviolet
divergence since $\chi(k)$ behaves as $k^{-2}$ for large $k$. A more
physical choice for ${1\over V}\langle|U_{\bf k}|^2\rangle$
would not yield such a divergence.

{}From result (12) it turns out that, without including
interaction terms among the quasiparticles, described by the anharmonic term
$H_1$,
the energies of the elementary
excitations are not affected by the random external field; in particular the
velocity of sound is still given by the Bogoliubov value
$c=(\rho_0v_0)^{1/2}/m$. In Section 4 we will calculate the
first correction to the velocity of sound as well as its damping due to the
random external potential.

\bigskip
\bigskip
\par\noindent
{\bf 3. SUPERFLUID DENSITY AND DEPLETION OF THE CONDENSATE}
\bigskip
The current density associated with the one-fluid hamiltonian (1) is
given by
$$
\hat{{\bf j}}({\bf r}) = {\partial E\over\partial{\bf v}_s} =
\rho_0\hat{{\bf v}}_s({\bf r})+\hat{\rho}'({\bf r})\hat{{\bf v}}_s({\bf r})
\eqno(14)
$$
where we have separeted the first and the second order terms.
By taking the Fourier transform of the current operator we obtain
$$
\hat{{\bf j}}_{\bf q}\;=\;i{\bf q}\;\rho_0\hat{\varphi}_{\bf
q}+{1\over\sqrt{V}}\sum_{\bf k}i{\bf k}\hat{\varphi}_{\bf k}
\hat{\rho}'_{{\bf q}-{\bf k}}
\eqno(15)
$$
The first term is longitudinal, whereas the second one contains both a
longitudinal as well as a transverse component. By inserting in eq. (15) the
expressions for the velocity potential and density (6) in terms of
the quasiparticle operators $\hat{c}_{\bf k}^\dagger,\hat{c}_{\bf k}$,
one can distinguish in a natural way between different contributions to the
current operator. The first term in (15) is coupled to single elementary
excitations: we call this term the one-phonon contribution to the current. The
second term can be either the product of one quasiparticle operator and of the
external field (hereafter referred to as disorder contribution) or the product
of two quasiparticle operators (two-phonon contribution).

Let us choose ${\bf q}$ in the $z$ direction: ${\bf q}=(0,0,q)$ and let us
consider the tranverse component of the current operator $\hat{j}_{\bf q}^x$.
In terms of the elementary excitation operators it is written as
$$
\hat{j}_{\bf q}^x\;=\;\hat{j}_{\bf q}^{x(1)}+\hat{j}_{\bf q}^{x(2)}\;=\;
-{1\over V}\sum_{\bf k} i\;k_x \left({\rho_0\omega_k\over2}\right)^{1/2}
{U_{{\bf q}-{\bf k}}|{\bf q}-{\bf k}|^2
\over\omega_{{\bf q}-{\bf k}}^2 k}
(\hat{c}_{\bf k}+\hat{c}_{-{\bf k}}^\dagger)\;+\;\hat{j}_{\bf q}^{x(2)}
\eqno(16)
$$
where the operator $\hat{j}_{\bf q}^{x(2)}$ arises from
two-phonon effects and does not depend upon the external field.

There is an important point to stress here.
The transverse current operator (arising from the second term in the r.h.s. of
eq. (15)) is fixed by the anharmonic term in the hamiltonian.
This implies that, for a calculation of the corresponding
matrix elements to the lowest order, we can safely
use the eigenstates and
eigenvalues of the harmonic hamiltonian $\hat{H}_0$. The situation is
different in the longitudinal channel, since the longitudinal component of
the current operator (15) contains a term, the one-phonon contribution,
which is zeroth order in the anharmonic perturbation. In the next Section we
will see that the anharmonic term $\hat{H}_1$ has to be explicitly included
in the
hamiltonian in order to study the longitudinal channel in the proper way.

The normal component of the fluid is
obtained by considering the static transverse current-current response which,
at $T=0$, can be written as
$$
\chi^T(q)\;=\;\sum_n
{|\langle0|\hat{j}_{\bf q}^x|n\rangle|^2\over\omega_n}
+{|\langle0|\hat{j}_{-{\bf q}}^x|n\rangle|^2\over\omega_n}\;\;\;.
\eqno(17)
$$
The normal density is then obtained from the long wavelength limit of
equation (17)$^9$
$$
\rho_n\;=\;\lim_{q\rightarrow0}\chi^T(q)
\eqno(18)
$$
while the  superfluid density is defined as $\rho_s = \rho_0 - \rho_n$.
In eq. (17) $|0\rangle$ is the ground state and the sum is performed over a
complete
set of excited states with energies $\omega_n$. If we now insert
the expression (16) for the current operator into eq. (17) we can distinguish
the
contribution to the response function arising from two-phonon effects from the
one fixed by disorder. The former can be shown to vanish in the
long wavelength limit as $q^2$. The latter
is easily obtained by a direct calculation of the matrix elements using the
diagonal hamiltonian (4).
The final result for the normal density is
given by
$$
\rho_n\;=\;\rho_0{1\over V}\sum_{\bf k}k_x^2{k^2\over\omega_k^4}
{1\over V} \langle|U_{\bf k}|^2\rangle
\;\;\;.
\eqno(19)
$$

The fact that the random external potential gives rise to a normal component in
the fluid can be also regarded as the consequence of the fact that
translational invariance is broken by the external field and therefore the
current is not conserved$^{10}$.

Result (19) needs some comments. First of all eq. (19) shows that the normal
component induced by the random external potential is not fixed just by the
long
wavelength properties of the system. In order to assure the convergence of the
integral in (19) the behaviour of the elementary excitation spectrum at high
momenta is important. A second point concerns the validity of the one-fluid
hamiltonian (1)  when a normal component is
present in the fluid.
Result (19) can be interpreted as the first term of the perturbative expansion
of the normal density in the external field $U$.
It is finally
worth noticing that in the case of an ideal Bose gas ($\omega_k=k^2/2m$)
the integral in eq. (19) diverges at small $k$'s revealing the instability of
the ideal Bose gas in the presence of disorder.

In the DBG limit the integral in (19) can be calculated analytically by
inserting the Bogoliubov value (7) for the energies $\omega_k$ and using the
white noise approximation for ${1\over V}\langle|U_{\bf k}|^2\rangle$. One gets
$$
\rho_n\;=\;{m^3\over6\pi^{3/2}}R_0\left({n\over a}\right)^{1/2}
\eqno(20)
$$
where $n$ is the particle density. The applicability of result (20)
is subject to the condition of weak disorder $\rho_n << \rho_0$ or,
equivalently,
$$
{m^2R_0 \over \sqrt{na}} << 1\;\;\;.
\eqno(21)
$$
 In 2D one finds
$$
\rho_n\;=\;{m^3\over8\pi^2}R_0{1\over a}\;\;\;.
\eqno(22)
$$
Both eqs.(20) and (22) are in agreement with the results of Refs. [3,11].

We now calculate the depletion of the condensate due to the random potential.
To this purpose let us write the macroscopic component of the particle field
operator $\hat{\Psi}({\bf r})$, in terms of density and phase fluctuations. The
macroscopic component of the particle operator is formally given by
$$
\hat{\Psi}_M({\bf r})=\sqrt{\hat{n}_0}e^{i\hat{\Phi}({\bf r})}
\eqno(23)
$$
where $\hat{n}_0$ is the operator of the number of particles in the
condensate and $\hat{\Phi}({\bf r})$ is the phase operator related to the
velocity
potential by the usual relation $\hat{\varphi}({\bf r})=\hat{\Phi}({\bf r})/m$.
In the DBG limit, where $n_0\simeq n=\rho_0/m$, the following expansion is
valid
up to first order in the density and phase fluctuations
$$
\hat{\Psi}_M({\bf r})\;=\;\sqrt{n}\left(1+{1\over2nm}\hat{\rho}'({\bf r})+
i\hat{\Phi}({\bf r})\right)
\eqno(24)
$$

By taking the Fourier transform of eq. (24) one finds
$$
\delta\hat{a}_{\bf q} = \hat{a}_{\bf q}-\langle\hat{a}_{\bf q}\rangle =
i\sqrt{n}\delta\hat{\Phi}_{\bf q}
+{1\over2\sqrt{n}m}\delta\hat{\rho}_{\bf q}
\eqno(25)
$$
with $\langle\hat{a}_{\bf q}\rangle=\sqrt{n}\delta_{{\bf
q},0}+1/(2\sqrt{n}m)\langle\hat{\rho}_{\bf q}'\rangle$ (we have used
$\langle\hat{\Phi}_{\bf q}\rangle=0$). The momentum distribution is thus
given by
$$
n(q)=\langle\delta\hat{a}_{\bf q}^{\dagger}\delta\hat{a}_{\bf q}\rangle +
|\langle\hat{a}_{\bf q}\rangle|^2=\;{\rho_0\over m}
\langle\delta\hat{\Phi}_{-{\bf q}}\delta\hat{\Phi}_{\bf q}\rangle +
{1\over 4m\rho_0}
\langle\delta\hat{\rho}_{-{\bf q}}\delta\hat{\rho}_{\bf q}\rangle -{1\over 2}
+ {1\over 4m\rho_0} |\langle\hat{\rho}_{\bf q}'\rangle|^2
\eqno(26)
$$

The first term in the right hand side of equation (26) represents
the contribution to the momentum distribution given by the phase
fluctuations. In the long wavelength limit this
term gives the main contribution and is responsible for the
well known $1/q$ divergence of $n(q)$$^{12}$. In the
region of higher momenta also the density fluctuations
$\langle\delta\hat{\rho}_{-{\bf q}}\delta\hat{\rho}_{\bf q}\rangle$
become important. The constant term $-1/2$ arises from the
commutation relation between density and phase operators.
It is worth noticing that the
random potential explicitly affects the result for the
momentum distribution because the
expectation value of the particle operator $\hat{a}_{\bf q}$ on the ground
state is no longer zero for ${\bf q} \ne 0$:
$$
\langle\hat{a}_{\bf q}\rangle =
{1\over 2\sqrt{n} m}\langle\hat{\rho}^{\prime}_{\bf q}\rangle
= -{1\over 2\sqrt{n} m}{1\over \sqrt{V}} U_{\bf q} \chi(q)
\eqno(27)
$$

By using  the Bogoliubov results for the density
and phase fluctuations (corrections to these quantities due to disorder
give rise to
higher order effects) we finally obtain
the result
$$
n(q)\;=\;{1\over2\omega_q}\left({q^2\over2m}+nv_0-\omega_q\right)
+{n\over4}{1\over V}\langle|U_{\bf q}|^2\rangle{q^4\over\omega_q^4}
\eqno(28)
$$
where the first term coincides with the prediction of
the Bogoliubov model.
The depletion is obtained by integrating result (28) and we find the usual
Bogoliubov value with a correction proportional to $\rho_n$:
$$
\Delta n\;=\;n-n_0\;=\;{1\over V}\sum_{{\bf q}\neq0}n(q)\;=\;{8\over3\pi^{1/2}}
(na)^{3/2}+{3\over4m}\rho_n
\eqno(29)
$$
Result (29) is in agreement with the findings of Ref. [3]. In particular it
shows that the relative depletion of the condensate due to disorder
is a factor $3/4$ smaller than the corresponding reduction of the
superfluid density. This result holds for any choice of the average
${1\over V}\langle|U_{\bf k}|^2\rangle$ since the same integral
over momenta is involved in the calculation of both  $\rho_n$ and $\Delta n$
(see eqs.(19) and (28)).
We stress however that eq. (29) holds only
in the presence of weak disorder. It cannot be  easily extrapolated
to the large external fields.

\bigskip
\par\noindent
{\bf 4. VELOCITY AND DAMPING OF SOUND}
\bigskip

In the present Section we discuss the effect of disorder on
the dispersion of the phonon mode at $T=0$.
To this purpose we first calculate the compressibility of the system,
$\chi(0)$, defined by the relation
$$
\chi^{-1}(0)={\partial^2\over\partial\rho^2}{E\over V}
\eqno(30)
$$
where $E$ is the ground state energy of the system. The contribution to $E$ due
to disorder is given by eq.(13) or equivalently, in the Bogoliubov limit,
by the first term in the r.h.s. of eq. (12).
After taking the second derivative
with respect to the density we find the result
$$
\chi^{-1}(0)=
\chi_0^{-1}(0)(1 +
\sum_{\bf k}{1\over V}\langle|U_{\bf k}|^2\rangle
{k^8\over 4\omega_k^6})
\eqno(31)
$$
where $\chi_0(0)$ is the compressibility of the DBG.
By making the white noise assumption (10) it is possible to relate result (31)
to the normal density $\rho_n$ given
by eq.(19). We find
$$
\chi^{-1}(0)=
\chi_0^{-1}(0)\left(1+{9\over4}{\rho_n\over\rho_0}\right)
\eqno(32)
$$
It is worth noticing that,
differently from the correction to the total energy as well as
to the first derivative $\partial E/\partial\rho$ (proportional to the
chemical potential), the contribution of disorder to the compressibility
is well defined also within the white noise assumption.
This is due to the fact that
the quantity
$\partial^2 \chi(k)/\partial\rho^2$, differently from $\chi$
and
$\partial \chi(k)/\partial\rho$
decreases as $k^{-4}$ when $k \to \infty$ and the corresponding integral
is well behaved for large $k$.

We are now ready to calculate the velocity of sound. Using the relation
$$
c^2={\rho_s\over\chi(0)}
\eqno(33)
$$
typical of superfluids and eqs. (20) and (32) for $\rho_n$ and
$\chi(0)$ we obtain the result
$$
c^2=c_0^2\left(1+{5\over4}{\rho_n\over\rho_0}\right)
\eqno(34)
$$
holding within the white noise assumption. Equation (34) shows
that the velocity of sound increases with disorder and that consequently
the phonon peak in the dynamic structure function
$$
S(q,\omega)=\int_{-\infty}^{+\infty}dt\;e^{i\omega t}\langle
[\delta\hat{\rho}_{-{\bf q}}(t),\delta\hat{\rho}_{\bf q}(0)]\rangle
\eqno(35)
$$
is shifted to the right according to the law
$$
S_{phonon}(q,\omega) = {\rho_sq\over2c}\delta(\omega-cq)\;\;\;.
\eqno(36)
$$
The normalization factor $\rho_sq/2c$ ensures that the
phonon exhausts the compressibility sum rule
$$
\chi(0) = 2lim_{_{q\to 0}} \int {1\over \omega}S(q,\omega)d\omega
\eqno(37)
$$
consistently with relation (33) for the sound velocity.

In the second part of the section we calculate the damping of phonons
due to collisions
with the external potential. To this purpose we calculate directly the
propagator defined by the time ordered product
$$
D_{11}(q,t)=-i\langle T(\hat{\rho}_{{\bf q}}'(t)\hat{\rho}_{-{\bf q}}'
(0))\rangle=-i\langle (\theta(t)\hat{\rho}_{{\bf q}}'(t)\hat{\rho}_{-{\bf q}}'
(0)\;+\;\theta(-t)\hat{\rho}_{-{\bf q}}'(0)\hat{\rho}_{\bf q}'(t))\rangle
\eqno(38)
$$
in terms of which the dynamic structure function is written as
$S(q,\omega)=-{1\over\pi}Im D_{11}(q,\omega)$.

As anticipated in Section 2 in order to calculate to the proper order the
effects of the external random potential in the longitudinal channel
it is essential to include the
anharmonic term (3) in the hamiltonian of the system.

We calculate the imaginary part of the propagator $D_{11}({\bf q},\omega)$
perturbatively in the interaction hamiltonian
$$
\hat{H}_{int}=\hat{H}_1+\hat{V}={1\over2\sqrt{V}}\sum_{{\bf k},{\bf k}'}
{\bf k}\cdot{\bf k}' \hat{\varphi}_{{\bf k}}\hat{\rho}_{{\bf k}'-{\bf k}}'
\hat{\varphi}_{-{\bf k}'}\;+\;{1\over\sqrt{V}}\sum_{{\bf k}}U_{\bf k}
\hat{\rho}_{-{\bf k}}'
\eqno(39)
$$
up to terms of second order in the random potential $\hat{V}$.
The calculation is straightforward and the details are given in the
Appendix. The relevant contribution to the imaginary part
of the propagator arises from the diagram of Figure 1. By adding this
contribution to the phonon peak (36) we find,
for small $q$, the result
$$
S(q,\omega)={\rho_sq\over2c}\delta(\omega-cq)\;+\;
{\omega^2q^2\over(\omega^2-c^2q^2)^2}{\rho_0\over2V}
\sum_{{\bf k}}{1\over V}\langle|U_{\bf k}|^2\rangle k_z^2{k^2\over\omega_k^3}
\delta(\omega-\omega_k)
\eqno(40)
$$
where the second term holds for $\omega$
not too close to $cq$: $|\omega-cq|>cq$.
In the hydrodynamic limit ($q\rightarrow0$, $\omega\rightarrow0$) result (40)
becomes
$$
S(q,\omega)={\rho_sq\over2c}\delta(\omega-cq)\;+\;
{q^2\omega^5\over(\omega^2-c^2q^2)^2}{\rho_0\over12\pi^2c^7}
{1\over V}\langle|U_{\bf k}|^2\rangle_{|\bf k|=\omega/c}
\eqno(41)
$$
Eq. (41) corresponds, in the limit of weak disorder (see eq.(21)), to the
first two
terms of the series of the most divergent diagrams characterizing the behaviour
of the propagator $D_{11}$, near the pole. The sum
of this series yields the following result for $S(q,\omega)$ near the pole
$$
S(q,\omega)={\rho_sq^2\over\pi}{2cq\Gamma(q)
\over(\omega^2-c^2q^2)^2+4c^2q^2\Gamma^2(q)}
\eqno(42)
$$
with the width $\Gamma(q)$ given by
$$
\Gamma(q)={1\over24\pi}{1\over c^3}q^4{1\over V}\langle|U_{\bf q}|^2\rangle
\;\;\;.
\eqno(43)
$$
Eqs. (42), (43) explicitly show that the phonon peak is broadened by the
presence of disorder.

It is not difficult to extend results (42), (43)
to the case of a strongly interacting Bose superfluid where quantum
hydrodynamics is expected to provide the proper description in the hydrodynamic
regime.
To this purpose we have to add to the anharmonic term (3) a contribution cubic
in the density fluctuations yielding, for $\hat{H}_1$, the more general
expression$^{13}$
$$
\hat{H}_1 = {1\over2\sqrt{V}}\sum_{{\bf k},{\bf k}'}
{\bf k}\cdot{\bf k}' \hat{\varphi}_{{\bf k}}\hat{\rho}_{{\bf k}'-{\bf k}}'
\hat{\varphi}_{-{\bf k}'}\;+\;
{1\over6\sqrt{V}}\left({d\over d\rho}{c^2\over\rho}\right)\sum_{{\bf k}{\bf
k}'}\hat{\rho}_{\bf k}'\hat{\rho}_{{\bf k}'-{\bf k}}'\hat{\rho}_{-{\bf k}'}'
\eqno(44)
$$
By accounting for the diagrams relevant for the calculation of the damping we
get the relevant result (see the Appendix for details)
$$
\Gamma(q)={1\over24\pi}{q^4\over c^3}{1\over V}\langle|U_{\bf q}|^2\rangle
\left(1+3{\rho_0^4\over c^4}({d\over d\rho}
{c^2\over\rho})^2\right)\;\;\;.
\eqno(45)
$$
In superfluid $^4$He the second term in eq. (45) turns out to be of the same
order as the first one.
Vice-versa in
the DBG  the derivative with respect to the density of $c^2/\rho$ is
of higher order in the scattering length and its contribution
can be consequently neglected.
Notice that the damping mechanism discussed above is dominant at small $q$
compared to the damping $\Gamma_{ph-ph}$ due to anharmonic interactions among
phonons, which exhibits a $q^5$ law (see Ref. [13]). In the dilute Bose gas and
using the white noise approximation we find the result
$$
{\Gamma(q)\over\Gamma_{ph-ph}(q)}={20\over3}{\rho_n\over\rho_0}{1\over qa}
\eqno(46)
$$

Starting from result (40) we can finally calculate the low $q$
behaviour of various energy moments of
the dynamic structure factor. These moments can be written as the sum of the
phonon contribution (first term in eq. (40)) and of the collisional term
(second term of eq. (40)). The latter contribution is always proportional to
$q^2$.
For example, by integrating  eq.(40) with respect to $\omega$, one finds the
following result for the low $q$ expansion of the non energy weigthed sum
rule (static structure
factor):
$$
S(q)=\int_0^\infty d\omega  S(q,\omega) = {\rho_sq\over2c} +
q^2{\rho_0 \over 2} \int d{\bf k}{1\over (2\pi)^3}
{1\over V}\langle|U_{\bf k}|^2\rangle {k_z^2 k^2\over\omega^5_k}
\eqno(47)
$$
By making the white noise approximation for
${1\over V}\langle|U_{\bf k}|^2\rangle$, one can relate the second term
to the normal density (19):
$$
S(q) = {\rho_sq\over2c} +
q^2 \rho_n {m\over 12\pi^2 na}
\eqno(48)
$$
Eq. (48) shows that the static structure factor, at low $q$, is exhausted by
the phonon peak. One immediately shows that the phonon peak exhausts also the
inverse energy
weighted sum-rule fixed by the compressibility of the system (eq. (37)).

For the energy weighted sum rule we find the result
$$
m_1=\int_0^\infty d\omega \omega S(q,\omega)
= \rho_s {q^2\over2} + \rho_n {q^2\over2}
= \rho_0{q^2\over2}
\eqno(49)
$$
which coincides, as expected,
with the model independent f-sum rule, proportional to the total density
of the system.  In this case the contribution of
the collisional term is crucial in order to satisfy
the sum rule also at low $q$.
This contribution
can be always expressed in terms of the normal density, independently of the
white noise assumption. It is interesting to
remark that a similar behaviour is exhibited by transverse
spin excitations (magnons) in antiferromagnets$^{14}$. Also in this case the
magnon  exhausts only  a fraction of the energy weighted sum rule, the
remaining
part being, in this case, exhausted by multimagnon excitations.
The one-magnon contribution to the energy weighted sum rule is proportional
to the spin stiffness coefficient which plays the role
of the superfluid density. This analogy between disordered
bosons  and antiferromagnets is due to the fact that in
both cases the current is not conserved due to the lack of translational
invariance.

The fact that the phonon peak does not exhaust the f-sum rule implies
that the Feynman approximation
$$
\omega_F(q) = {\rho_0q^2\over 2S(q)}
\eqno(50)
$$
for the energy of elementary excitations does not coincide
with the phonon dispersion at low $q$. Actually using result (47)
for $S(q)$ one finds $\omega_F(q) \to cq\rho_0/\rho_s$ a value higher than
$cq$. It is also worth noticing that the Feynman ratio (50)
is affected by the $q^2$ correction in $S(q)$. Such a correction
is absent in translationally invariant systems.

\bigskip
\bigskip

\par\noindent ACKNOWLEDGMENTS

We wish to thank K. Huang for useful discussions. L.P. likes to thank the
hospitality of the Department of Physics of the University of Trento.

\vfill\eject

\par\noindent
{\bf APPENDIX}
\bigskip

In this appendix we evaluate the imaginary part of the
density-density propagator
$D_{11}({\bf q},\omega)$ in the long wavelength limit, to first
 order in $R_0$. We use
time dependent perturbation theory with the perturbative field
given in equation (39).

We need the free field propagators:
$$
iD_{11}^0(q,\omega)=\int_{-\infty}^{+\infty}dt \;e^{i\omega t}
\langle T(\hat{\rho}_{{\bf q}}'(t)\hat{\rho}_{-{\bf q}}'(0))\rangle_0 =
i {\rho_0q^2\over\omega^2-\omega_q^2+i0}\;\;\;\;\;
\eqno(A1)
$$
$$
iD_{22}^0(q,\omega)=\int_{-\infty}^{+\infty}dt \;e^{i\omega t}
\langle T(\hat{\varphi}_{{\bf q}}(t)\hat{\varphi}_{-{\bf q}}(0))\rangle_0 =
i{\omega_q^2\over\rho_0q^2}{1\over\omega^2-\omega_q^2+i0}
\eqno(A2)
$$
$$
iD_{12}^0(q,\omega)=\int_{-\infty}^{+\infty}dt \;e^{i\omega t}
\langle T(\hat{\rho}_{{\bf q}}'(t)\hat{\varphi}_{-{\bf q}}(0))\rangle_0 =
-{\omega\over\omega^2-\omega_q^2+i0}\;\;\;\;\;
\eqno(A3)
$$
where the subscript after the brackets $\langle...\rangle_0$ means that the
expectation value is taken on the ground state relative to the
unperturbed hamiltonian $H_0$.

The density-density propagator is obtained from the general formula
of perturbation theory
$$
iD_{11}({\bf q},t)={1\over\langle\hat{S}\rangle}\langle T(\hat{\rho}_{\bf q}'
(t)\hat{\rho}_{-{\bf q}}'(0)\hat{S})\rangle
\eqno(A4)
$$
where the time evolution operator $\hat{S}$ is defined by the series expansion
$$
\hat{S}=\sum_{n=0}^\infty {(-i)^n\over n!}\int_{-\infty}^{+\infty}dt_1\;...\;
\int_{-\infty}^{+\infty}dt_n \;T(\hat{H}_{int}(t_1)...\hat{H}_{int}(t_n))
\eqno(A5)
$$

It is better to use diagrams to represent the different terms arising from
equation (A4) after using Wick's theorem for time ordered products. The
anharmonic potential $\hat{H}_1$ is represented by a vertex
with three lines, whereas to the external random potential corresponds
a vertex with just one line.

At finite frequencies the leading correction to the imaginary part of
$D_{11}({\bf q},\omega)$ comes from terms which are
fourth order in the interaction hamiltonian $\hat{H}_{int}$.
These terms correspond to
diagrams containing two vertices relative to the anharmonic potential and two
random potential vertices.

It turns out that only one diagram is relevant and this is given in figure 1,
where we have represented $D_{11}^0$ by a full line, $D_{22}^0$ by a dashed
line and $D_{12}^0$ by a long-dashed line.
In fact the external random field is time independent. This means that only the
density-density propagator, which has a zero frequency component, can be
connected to the random potential vertices.
The contribution arising from the diagram in figure 1 to the density-density
propagator is given by
$$
D_{11}({\bf q},\omega)=D_{11}^0(q,\omega)\;+\;
{\omega^2\over\rho_0^2q^2}(D_{11}^0(q,\omega))^2
$$
$$
\;\;\;\;\;\;\;\;\;\;\;
\times\;
{1\over V}\sum_{{\bf k}}{1\over V}\langle|U_{{\bf k}+{\bf q}}|^2\rangle
k_z^2\;D_{22}^0(k,\omega)
(D_{11}^0(|{\bf k}+{\bf q}|,0))^2
\eqno(A6)
$$
The square of the free propagator $D_{11}^0$ in equation (A6) can be
approximated, for $\omega$ not too close to the pole
($|\omega-\omega_q|>\omega_q$), with the expression
$$
(D_{11}^0(q,\omega))^2 = {\rho_0^2q^4\over(\omega^2-\omega_q^2)^2}
\eqno(A7)
$$
Result (40) for the dynamic structure factor follows directly from
equations (A6) by using the relation
$S({\bf q},\omega)=-{1\over\pi}Im D_{11}({\bf q},\omega)$ and after taking
the $q\rightarrow0$ limit.

The imaginary part of the density-density self energy can be directly
evaluated from equation (A6). In the long wavelength limit one gets
$$
-Im \Sigma_{11}(q,\omega)= {\pi\over2}{\omega^2\over\rho_0q^2}{1\over V}
\sum_{\bf k}{1\over V}\langle|U_{\bf k}|^2\rangle k_z^2{k^2\over\omega_k^3}
\delta(\omega-\omega_k)
\eqno(A8)
$$
Result (43) for the damping in the phonon region follows from eq. (A8)
by using the relation
$$
\Gamma(q)= {\rho_0q\over 2c}(-Im\Sigma_{11}(q,\omega=cq))
\eqno(A9)
$$

In the case of strongly correlated systems the damping of the phonon mode
can be obtained in the long wavelength limit by using the anharmonic term (44).
In this case another diagram is relevant to the calculation of the damping and
it is given in figure 2.
The imaginary part of the self energy is given by
$$
-Im \Sigma_{11}(q,\omega)= {\pi\over2}{1\over\rho_0}{1\over V}
\sum_{\bf k}{1\over V}\langle|U_{\bf k}|^2\rangle
\left(k_z^2{\omega^2k^2\over\omega_k^3q^2}+\rho_0^4
({d\over d\rho}{c^2\over\rho})^2{k^6\over\omega_k^5}\right)
\delta(\omega-\omega_k)
\eqno(A10)
$$
and result (45) follows directly from relation (A9).

The present diagrammatic technique could be used for the calculation of
higher order corrections in the parameter $R_0$. To perform the
corresponding averaging one must
assume gaussian statistics for the random external field.

\vfill\eject

\par\noindent REFERENCES

$^1$ M. Ma, B.I. Halperin and P.A. Lee, Phys. Rev. B, {\bf 34}, 3136 (1986).

$^2$ M.P.A. Fisher, P.B. Weichman, G. Grinstein and D.S. Fisher, Phys. Rev. B

{\bf 40}, 546 (1989).

$^3$ K. Huang and H.F. Meng, Phys. Rev. Lett. {\bf 69}, 644 (1992).

$^4$ W. Krauth, N. Trivedi and D. Ceperley, Phys. Rev. Lett. {\bf 67}, 2307
     (1991).

$^5$ M. Makivic, N. Trivedi and S. Ullah, Phys. Rev. Lett. {\bf 71}, 2307
     (1993).

$^6$ L. Zhang, Phys. Rev. B {\bf 47}, 14364 (1993).

$^7$ P. Nisamaneephong, L. Zhang and M. Ma, Phys. Rev. Lett. {\bf 71}, 3830
     (1993).

$^8$ L.P. Pitaevskii, J. Exp. Theoret. Phys. (U.S.S.R.) {\bf 31}, 536 (1956).

$^9$ G. Baym, in {\it Mathematical Methods in Solid State and Superfluid
     Theory},

     R.C.Clark and G.H.Derrick eds. (Oliver and Boyd, Edinburgh, 1969),p.151.

$^{10}$ S. Stringari, in {\it Bose Einstein Condensation}, A.Griffin, D.Snoke

        and S.Stringari eds. (Cambridge University Press, in press).

$^{11}$ H.F. Meng, Ph. D. thesis, MIT (unpublished).

$^{12}$ T. Gavoret and Ph. Nozieres, Ann. Phys. (N.Y.) {\bf 28},349 (1964).

$^{13}$ E.M. Lifshitz and L.P. Pitaevskii, {\it Statistical Physics}

(Pergamon,Oxford, 1980), Part 2, pp. 98 and 134.

$^{14}$ S. Stringari, (Phys. Rev. B. in press).

\vfill\eject

\bigskip
\par\noindent FIGURE CAPTIONS

\bigskip

\par\noindent
Fig.1. The relevant diagram of the fourth order in $\hat{H}_{int}$ (eq. (39)).
The full line represents $D_{11}^0$, the dashed line $D_{22}^0$ and the
long-dashed line $D_{12}^0$.

\par\noindent
Fig.2. The other relevant diagram of the fourth order in $\hat{H}_{int}=
\hat{H}_1+\hat{V}$ with $\hat{H}_1$ given by eq. (44). The full line represents
$D_{11}^0$.

\bye